%Paper: hep-ph/9304300
%From: PHADS%TWNAS886.BITNET@pucc.Princeton.EDU
%Date: Tue, 27 Apr 1993 11:36 PDT
%Date (revised): Fri, 16 Jul 1993 10:15 PDT

\magnification\magstep1
\hsize5.45truein
\hoffset.8truein
\baselineskip24truept
\topskip10truept
{
\nopagenumbers
\hfill IP-ASTP-11-93

\hfill revised July, 1993

\null
\vfill

\centerline{\bf The Motion of Massive Test Particles in Dark Matter}
\centerline{\bf with an $a_0/r^2$ Energy Density}
\vskip.5in

\centerline{\it Achilles D. Speliotopoulos}
\footnote{}{\rm Bitnet address: PHADS@TWNAS886}

\vskip.5cm

\centerline{\it Institute of Physics}
\centerline{\it Academia Sinica}
\centerline{\it Nankang, Taipei, Taiwan 11529}

\vfill

\centerline{\bf Abstract}

\vskip.25in

{
 \baselineskip26truept
 \leftskip2in
 \rightskip2in
 \noindent The motion of massive test particles in dark matter is
 studied. It is shown that if the energy density of the dark matter
 making up a galactic halo has a large $r$ behavior of $1/r^2$, then
 contrary to intuition the motion of these test particles are not
 govern by Newtonian gravity, but rather by the equations of
 geodesic motion from Einstein's theory of general relativity.
 Moreover, the rotational velocity curves of orbiting massive test
 particles in this energy density do not approach a constant value at
 large $r$ but will instead always increase with the
 radius of the orbit $r_c$.
}
\vskip1truecm
\vfill
\supereject
}
\pageno=2

\noindent{\bf \S 1. Introduction}

One of outstanding problems in astrophysics today is to explain the motion
of bodies with orbits in the galactic halo. From Newtonian
dynamics one would expect that for a body in a circular orbit
$$
{v^2_\phi\over r_c}= {GM_g\over r_c^2}
$$
where $G$ is the gravitational constant, $M_g$ is the luminous mass of the
galaxy, $r_c$ is the radius of the circular orbit and $v_\phi$ is its
rotational velocity. If one were then to plot $v_\phi$ verses $r$ for
various bodies orbiting in a galactic halo one would expect to
obtain a rotational velocity curve which {\it decreases\/} as
$1/\sqrt r_c$. This does not, in fact, happen. Experimentally, it is
instead found that the rotational velocity curve approaches a constant
value at large $r$. To explain this result, the presence of ``dark
matter'', matter which has yet to be detected, has traditionally been
proposed. (See $[1]-[3]$ and the references contained therein.)
Namely, it has been postulated that the galactic halo is filled with
a gas of weekly interacting particles with a mass density $\rho$.
With their presence the total mass contained within the radius of the
orbiting body changes and the Newtonian equation of motion now becomes
$$
v_\phi^2 = G\left({M_g\over r_c} + {4\pi\over r_c} \int_0^{r_c}\rho
r^2dr\right).
$$
If we now take $\rho \approx a_0c^2/(Gr^2)$ for large $r$ where
$a_0$ is a dimensionless constant, then as long as
$$
a_0 \gg {M_gG\over 4\pi R_g c^2}\>,
\eqno(1)
$$
the $M_g$ term may be neglected and a constant $v_\phi$ with value
$$
{v_\phi^2\over c^2} \approx {4\pi a_0}\>,
$$
can be obtained. ($R_g$ is the point at which the velocity curves
become a constant and is identified as the radius of the luminous
galaxy.) Explanation of the rotational velocity curves then reduces to
finding and detecting candidates for this dark matter which will
have the correct large $r$ behavior.

The basic premise of this argument is that Newtonian dynamics and
gravity will still be valid even after the introduction of a
$1/r^2$ energy density into the system. This need not be true. One
should remember that Newtonian gravity is only an approximation of
Einstein's theory of general relativity and that the very act of
introducing a mass density $\rho\sim1/r^2$ for dark matter
introduces an  {\it unconfined\/} energy density into the system. Its
presence cannot help but have an affect on the geometry of the spacetime
in the halo, and, consequently, on the motion of bodies orbiting
there.

In this paper we shall show that the standard argument using
Newtonian dynamics and gravity for the existence of dark matter with
a $1/r^2$ energy density is inconsistent. The
introduction of an energy density which behaves as $1/r^2$ for
large $r$ changes the geometry of the spacetime so drastically that
the motion of bodies in the galactic halo is necessarily
non-Newtonian. The basic premise that Newtonian gravity is still
valid even after the introduction of this dark matter is incorrect.
Furthermore, after using the full geodesic equation from Einstein's
theory of general relativity to analyze circular orbits in the
$1/r^2$ energy density, we find that $v^2_\phi \sim q r_c^q$ for
$0<q\le1$. The rotational velocity always {\it increases\/} with
$r_c$ and will not approach a constant value. In fact, for a static,
spherical geometry there are only a very narrow range of essentially
unphysical choices for the energy density which will
give a constant $v_\phi$ for $r>R_g$.

\noindent{\bf \S 2. Geometry of the $a_0/r^2$ energy density}

We begin by modeling the galaxy as a sphere of mass $M_g$ and radius
$R_g$ which is surrounded by a galactic halo made up of dark matter.
The most general static, spherically symmetric metric is known to be
$[4]${}
$$
ds^2 = -f(r) dt^2 + h(r)dr^2 + r^2(d\theta^2 + \sin^2\theta d\phi^2)\>,
$$
where $f$ and $h$ are unknown functions of $r$ only which need to be
determined. As usual, we write the energy momentum tensor for the system as
$$
T_{\mu\nu} = \rho u_\mu u_\nu + p (g_{\mu\nu} + u_\mu u_\nu)\>,
$$
where $\rho$ is the energy density of the particles in the halo, $p$
is their pressure, and $u_\mu$ is an unit velocity vector in the
direction of the timelike Killing vector for the system. From
Einstein's equations $[4]$ we then obtain
$$
\eqalignno{
8\pi\rho =
         &
         {1\over r^2} {d\>\>\> \over
         dr}\left\{r\left(1-{1\over h}\right)\right\} \>,
         \cr
8\pi p =
        &
        {f'\over rhf} - {1\over r^2} \left( 1-{1\over h}\right)\>,
        \cr
8\pi p =
        &
        {1\over 2} {f'' \over fh} - {1\over 4} {f'\over fh}
        \left({f'\over f}+{h'\over h}\right) + {f'\over 2rfh} -
        {h'\over 2rh^2}\>,
        &(2)
        \cr
}
$$
where we are using units in which $G=c=1$ and the primes denote derivatives
with respect to $r$.

Suppose now that the energy density of dark matter is $\rho
\approx a_0/r^2$ for $r>R_g$ where $a_0>0$ is a (dimensionless)
constant. Then the first equation in $(2)$ is trivial to
integrate giving
$$
h^{-1} = 1-8\pi a_0 -{K\over r}\>,
$$
where $K$ is an integration constant. Since in the absence of dark
matter we would expect to have obtained the Schwarzchild solution, we
identify $K$ with $2M_g$. Next, notice that as long as the Newtonian
bound $(1)$ on $a_0$ holds
$$
a_0 \gg {M_g\over 4\pi r}\>,
$$
and we can neglect this term in $h$ and can approximate $h^{-1}
\approx 1-8\pi a_0$ as a constant for $r>R_g$. We shall justify this
approximation later.

The difference of the second two equations gives
$$
0 = {1\over 2} {f'' \over f} - {1\over 4} {f'\over f}
        \left({f'\over f}+{h'\over h}\right) - {1\over 2r}\left({f'\over f}+
        {h'\over h}\right) +{(h-1) \over r^2}.
\eqno(3)
$$
Since $h$ is a constant for $r>R_g$, it is straightforward to solve,
$$
f(r) \approx \left( k_{+} r^{q_{+}/2} + k_{-} r^{q_{-}/2}\right)^2\>,
$$
where
$$
q_{\pm} = 2\pm2\left(1-16\pi a_0\over 1-8\pi a_0\right)^{1/2}\>,
$$
and $k_\pm$ are integration constants. Because $f>0$, $a_0\le
1/16\pi$ otherwise $f$ will contain oscillatory solutions. Since we
are working in the large $r$ limit, and since $q_{+}>q_{-}$, only one
of the two solutions to $(2)$ will survive. Linear
combinations of the two will not. We can thus consider each solution
independently and for convenience we shall write $f_{\pm}= k_{\pm}
r^{q_{\pm}}$. Then
$$
8\pi p_{\pm} \approx {q_{\pm}+1-h\over hr^2}\>,
$$
and we see that the pressure also varies as $1/r^2$. Moreover,
$$
p =\left(q_\pm\over 4-q_\pm\right)\rho\>,
$$
and since $q_{+}\ge 2$, $p_{+} > \rho$. Because $p\le \rho/3$, we must
therefore exclude the $(+)$ solutions as being unphysical. The only
physical solutions are the $(-)$ solutions, and we thus set $q =
q_{-}$, $p=p_{-}$ and $f=kr^q$. Furthermore, $0< q\le1$
while $M_g/(4\pi R_g)< a_0\le 3/(56\pi)$. Consequently, $1< h \le
7/4$, and, since $f\sim r^q$, we can see explicitly that spacetime in
the presence of the $1/r^2$ energy density is not flat, but is
instead quite curved.

We next consider the motion of a massive test particle in a static,
spherically symmetric geometry with a velocity $v_\mu$ such that
$-1=v_\mu v^\mu$. This constraint gives
$$
-1 = -f\left(dt\over d\tau\right)^2 + h \left(dr\over d\tau\right)^2 + r^2
\left(d\theta\over d\tau\right)^2 + r^2\sin^2\theta \left(d\phi\over
d\tau\right)^2 \>,
$$
where $\tau$ is the proper time of the particle. Working in the
equatorial $\theta = \pi/2$ plane, and using energy and angular momentum
conservation,
$$
E = f {dt\over d\tau}\qquad,\qquad L = r^2{d\phi\over d\tau} \>,
\eqno(4)
$$
where $E$ and $L$ are the energy and orbital angular momentum per unit mass,
respectively, of the particle, we obtain
$$
0=\left(dr\over dt\right)^2 + V(r)\>,
$$
where
$$
V(r) = {f^2\over E^2 h}\left(1+{L^2\over r^2}\right) - {f\over h}\>,
$$
is an effective potential energy. For circular motion, $r=r_c$, the
radius of the circular orbit which is a constant in time.
Consequently, $V(r_c)=0$ and $V'(r_c)=0$, giving
$$
E^2 = {2f^2(r_c)\over 2f(r_c) - r_cf'(r_c)}\qquad,\qquad
{L^2 \over r_c^2} = {r_cf'(r_c)\over 2f(r_c) - r_c f'(r_c)}.
\eqno(5)
$$
Defining the rotational velocity as
$$
v_\phi = r {d\phi\over dt}
$$
then from $(4)$ and $(5)$ we find that for circular motion,
$$
v_\phi^2 = {1\over 2}r_cf'(r_c)\>.
\eqno(6)
$$
This equation holds for any $f$. In particular, if we are dealing with a
spherical mass $M$ in free space, then $f=(1-2M/r)$ and $(6)$ reduces
to what one obtains from Newtonian gravity. If, on the other hand,
$\rho\sim 1/r^2$, then from the above $f=kr^q$, so that
$$
v_\phi^2 = {1\over 2}qkr_c^q.
$$
For $q\ne 0$, $v_\phi$ always {\it increases\/} with the radius of
the orbit and never approaches a constant value as one would naively expect
from Newtonian gravity.

In the above solution of $(3)$ for $f$ we have neglected the
contribution of $h'/h$ in comparison to $f'/f$. We shall now
justify this approximation. First, we note that for the galaxy,
$2M/r\ll 1$ and, due to the bound $(1)$ on $a_0$, this term is very
small in comparison to $1-8\pi a_0$. Next, note that while $h'/h \sim
2M/r^2$, the solution we obtained by taking $h$ as a constant gives
$f'/f\sim 1/r$. Since $r$ is large, we would expect $h'/h$ to have
a very small affect on the $kr^q$ solution of $(3)$. Consequently, we
can take $h'/h$ as a small perturbation and solve $(3)$
perturbatively about the $kr^q$ solution. After doing so, we find
that to first order in $2M/r$,
$$
f = kr^q\left[1 - {2Mh_0\over (3-q)r}\left({q\over 2} +
2h_0+1\right)\right]\>,
$$
where $h_0^{-1} \equiv 1-8\pi a_0$. The inclusion of the $2M/r$ term
in $h$ modifies the $kr^q$ solution very slightly since $2M/r\ll 1$,
and $1<h_0\le 7/4$. Consequently, we were justified in neglecting this
contribution to $f$.

\noindent{\bf \S 3. Constant $v_\phi$ energy density}

We now ask whether or not it is possible for any
physically reasonable $\rho$ to result in a $v_\phi$ which will be
constant for $r>R_g$. Using $(6)$, we find that for $v_\phi$ is to be a
constant outside of the galaxy, $f$ must then have the
approximate form of $f_v\approx 2v^2_\phi\log(r/r_0)$ for $r$ greater
than some $r_0$. (The subscript $v$ will denote the fact that we are
looking for solutions of Einstein's equations which will result in a
constant $v_\phi$.) Because $f_v>0$, $r>r_0$, and since $v_\phi$ is a
constant only outside of the galaxy, we shall identify $r_0$ with
$R_g$. Since $f_v$ is now given and $h_v$ unknown, $(3)$ becomes a
differential equation for $h_v$ which may be written as
$$
0={d\>\>\over dy}\left(1\over h_v\right) - {1+2y\over y}\left(1\over
h_v\right) + {4y\over 1+2y} \>,
$$
where $y = \log(r/R_g)>0$. Its solution is straightforward,
$$
{1\over h_v} = y e^{2y} \left(c_h + 2\int^\infty_{2y} {e^{-t}\over 1+t}
dt\right)\>,
$$
where $c_h$ is an integration constant. Then from $(2)$ we find that:
$$
\eqalignno{
8\pi r^2\rho_v =
                &
                1 + {4y\over 1+2y} - (1+3y)e^{2y}\left(c_h +
                2\int^\infty_{2y} {e^{-t}\over 1+t}dt\right)\>,
                \cr
8\pi r^2 p_v =
                &
                (1+y) e^{2y} \left(c_h +2\int^\infty_{2y} {e^{-t}\over
                1+t}dt\right) - 1.
                &(7)
                \cr
}
$$
Physically, $\rho_v\ge 3p_v$. This gives an upper bound of $0.0013$
for $c_h$. As $h>0$ for all $y>0$, $c_h\ge0$. Consequently, $0\le
c_h\le 0.0013$ and there are only a very narrow range of values for
$c_h$ which will result in a physically reasonable $\rho_v$ and
$p_v$. Moreover, if $c_h>0$, then it is only when $y$ is between some
$y_{min}\>$ and $y_{max}\>$ that $\rho_v\ge 3p_v$. Outside of these two
bounds the energy density must have a different form and $v_\phi$
cannot be a constant. If, on the other hand, $c_h=0$ then
$\rho_v \ge 3p_v$ for all $y>0.627$. As long as
$r>e^{0.627}R_g$ the energy density given in $(7)$ for $c_h=0$ will
result in a constant rotational velocity curve outside of the galaxy.
For $r<e^{0.627}R_g$ the energy density will have a different form
and $v_\phi$ will not be a constant, as expected.

\noindent{\bf \S 4. Concluding Remarks}

We have thus shown that if the energy density $\rho$ of dark matter
behaves as $1/r^2$ for large $r$, then
$f \approx k r^q$ for some constant $k$ while $h\approx 1-8\pi a_0$.
Consequently, the premise that $\rho\sim1/r^2$ for large
$r$ contradicts the premise that Newtonian gravity is valid in the halo.
The introduction of this energy density, which is not confined but
spread over a large area, alters the geometry of spacetime so
drastically that Newtonian dynamics is no longer valid in the halo.
The assumption that Newtonian gravity is valid
even after the introduction of dark matter with a $1/r^2$ energy
density is incorrect. In fact, contrary to what is expected from
Newtonian gravity, circular orbits for this energy density have a
$v_\phi^2 = qkr_c^q/2$ which always increases with $r_c$. It
never approaches a constant, although for very small $q$ it increases
very slowly.

{}From a physicist point of view, however, the major problem with
using a $1/r^2$ energy density for dark matter is not that there was
an inconsistency in the argument for its introduction since this is
straightforwardly resolved by using general relativity instead of
Newtonian gravity to analyze the system; nor is it that $v_\phi$
always increases with $r_c$ since $q$ may be taken to be quite small
so that any increase in $v_\phi$ occurs very gradually
(with one cavet; see $[5]$). It is rather that the spacetime in a
$1/r^2$ energy density is so curved that Newtonian dynamics is no
longer valid. If the energy density of the dark matter in the
galactic halo truly does behave as $1/r^2$ for large $r$, this
will present great difficulties in interpreting extragalactic
astronomical observations. As most galaxies have a halo, including
presumeably our own, light from a distant galaxy
would then first have to pass through its own halo, a region of
curved spacetime, and then our through own galactic halo, another
region of curved spacetime, before we can observe it. Since the usual
assumption is that light from other galaxies passes through
a spacetime which is essentially flat before it reaches us, if the
energy density of dark matter has a $1/r^2$ large $r$ behavior, then
all of the extragalactic observational data would have to be
re-evaluated and interpreted. This is just one of the many
problems that would arise from using a $1/r^2$ energy density for
dark matter which have yet to be addressed.

The $1/r^2$ energy density is an unconfined energy density which,
presumeably, extends for large distances into the spacetime. As the
universe is known to be expanding and thus changing with time, one
may question the validity of using a static solution of Einstein's
equations to analyze the motion of test particles in the dark matter
as we have done. As, however, the galactic rotation curves only
extend out to a few galactic radii, we are only interested in the
behavior of the motion of bodies relatively close to the galaxy and
in this region a static approximation is certainly valid. Of course,
at some $r\gg R_g$ the $1/r^2$ energy density must be cutoff-ed and
our analysis will no longer be valid, but this will only happen
outside the region we are interested in. We should also note that
precisely the same static approximation is made in the standard
analysis of the motion test particles in the galactic halo using
Newtonian gravity. The important point here is not whether the
use of a static, $1/r^2$ energy density to model the galactic halo is
a valid approximation or not, but rather that our analysis of the
motion of test particles in a $1/r^2$ energy density using general
relativity holds in precisely the same regime in which the standard
analysis using Newtonian gravity was {\it presumed\/} to be valid.

As $(3)$ is a non-linear second order differential equation in
$f$, it may be that we truly cannot neglect the
$h'/h$ term in $(3)$ no matter how small $2M/r$ is. In $\bf\S 3$,
however, we have shown that in a static, spherical
geometry a rotational velocity curve which is truly a constant for
$r>R_g$ is obtainable only for the very special choices of the $\rho_v$
given in $(7)$. The only approximation made in obtaining $(7)$ was
once again that the system is static, and spherically symmetric.
Notice, however, that if $(7)$ is truly the energy density of dark
matter, this would mean that no matter what the internal
properties of the galaxy are, once one leaves it the energy density
of the particles making up its' halo has to be determined within one
part in $1000$. This is extraordinarily and prohibitively
restrictive. The form that $\rho_v$ takes is very particular and it
is difficult to imagine a physical process  which will not only
reproduce it, but also determine $\rho_v$ to such a high degree of
accuracy. On this basis alone we would tend to rule out $\rho_v$ as a
physically viable energy density for dark matter. We also note,
however, that because $f\sim \log(r/R_g)$, once again the spacetime
with this energy density is curved and in using $\rho_v$ as the
energy density of dark matter we would once again be faced with the
problem of interpreting the observational data.

The analysis done in this paper was done for a very
special system under some very restrictive conditions. For example,
although we have used a spherically symmetric geometry to model the
galaxy, most observed galaxies are axisymmetric and have a definite
angular velocity. It is, moreover, not even clear whether the
experimental data gives rotational velocity curves which are truly a
constant or whether they are instead slightly increasing or
decreasing with $r_c$. All that we are comfortable concluding from
this analysis, therefore, is that one must be much more careful about
introducing any unconfined energy density for dark matter into the
system. Not only  must it be able to explain the experimental
rotational velocity curves {\it within the framework of general
relativity}, but one must also consider the subsequent affects of
this energy density on the geometry of the spacetime. As we have
seen, for the $1/r^2$ energy density these affects are considerable.
\vfill
\supereject
\centerline{\bf Acknowledgements}

ADS would like to thank K.-W. Ng for many helpful discussions while
this paper was being written. This work is supported by the National
Science Council of the Republic of China under contract number NSC
82-0208-M-001-086.
\vfill
\centerline{\bf REFERENCES}
\item{$[1]$}V. C. Rubin, W. K. Ford, and N. Thonnard, {\sl Ap. J.
            Letters}, {\bf 225} L107 (1978).
\par
\item{$[2]$}V. Trimble, {\sl Ann. Rev. Astron. and Astrophys.}, {\bf
            25} 425 (1987).
\par
\item{$[3]$}E. W. Kolb and M. S. Turner, {\sl The Early Universe},
            Chapter 1 (Addison-Wesley Publishing Company, Inc.,
            New York, 1990). E. W. Kolb and M. S. Turner, {\sl The
            Early Universe: Reprints}, Chapter 1 (Addison-Wesley
            Publishing Company, Inc., New York, 1988).
\par
\item{$[4]$}R. M. Wald, {\sl General Relativity}, Chapter 6 (The
                University of Chicago Press, Chicago, 1984).
\par
\item{$[5]$}Since we have taken $a_0\gg M_g G/(4\pi R_gc^2)$,
there is a lower bound below which we would not be able to decrease
$q$. For a typical galaxy with $M_g\approx 10^{12}M_\odot$ and
$R_g\approx 5$ kpc, this bound turns out to be $\sim 2\times
10^{-5}$. Note also that because $p = q/(4-q)\rho$, very small $q$
would correspond to very ``cold'' dark matter which would have a
temperature which is very much smaller than its mass.
\end